\documentclass[graybox]{svmult}

\usepackage{type1cm}        
\usepackage{makeidx}         
\usepackage{graphicx}        
\usepackage{color}
\usepackage{epstopdf}
\usepackage{multicol}        
\usepackage[bottom]{footmisc}

\usepackage{newtxtext}       %
\usepackage[varvw]{newtxmath}       
\usepackage{subcaption}
\usepackage{subcaption}

\makeindex             

\begin{document}

\title*{Wetting simulation of the porous structure of a heat pipe using  an eXtended Discontinuous Galerkin Method and a Parameterized Level-Set}

\titlerunning{Wetting simulation using a Parameterized Level-Set}
\author{Irina Shishkina\orcidID{0000-0002-1094-8386 } 
and\\ Matthias Rieckmann\orcidID{0000-0003-2199-9573}
and\\ Martin Oberlack\orcidID{0000-0002-5849-3755}
and\\ Florian Kummer\orcidID{0000-0002-2827-7576}}
\institute{Irina Shishkina \at TU Darmstadt, Otto-Berndt-Str. 2, 64287 Darmstadt, Germany, \email{shishkina@fdy.tu-darmstadt.de}
\and Matthias Rieckmann \at TU Darmstadt, Otto-Berndt-Str. 2, 64287 Darmstadt, Germany, \email{rieckmann@fdy.tu-darmstadt.de}
\and Martin Oberlack \at TU Darmstadt, Otto-Berndt-Str. 2, 64287 Darmstadt, Germany, \email{oberlack@fdy.tu-darmstadt.de}
\and Florian Kummer \at TU Darmstadt, Otto-Berndt-Str. 2, 64287 Darmstadt, Germany, \email{kummer@fdy.tu-darmstadt.de}}
%
%
\maketitle

\abstract{We perform high-order simulations of two-phase flows in capillaries, with and without evaporation. Since a sharp-interface model is used, singularities can arise at the three-phase contact line, where the fluid-fluid interface interacts with the capillary wall. These singularities are especially challenging when a highly accurate, high-order method with very little numerical diffusion is used for the flow solver. In this work, we employ the eXtended Discontinuous Galerkin (XDG) method, which has a very high accuracy but a severe limit regarding e.g., the time-step restriction. \\
To address this challenge and enhance the stability of our numerical method we introduce a novel approach for representing a moving interface in the case of two-phase flows. We propose a global analytical representation of the interface-describing level-set field, defined by a small set of time-dependent parameters. Noteworthy for its simplicity and efficiency, this method effectively addresses the inherent complexity of two-phase flow problems. Furthermore, it significantly improves numerical stability and enables the use of larger time steps, ensuring both reliability and computational efficiency in our
simulations. \\
We compare different analytic expressions for level-set representation, including the elliptic function and the fourth-order polynomial, and validate the method against established literature data for capillary rise, both with and without evaporation. These results highlight the effectiveness of our approach in resolving complex interfacial dynamics.
}


\section{Introduction}
\label{sec:1}
Capillaries play a crucial role in various technological applications due to their unique properties in fluid dynamics. In medical technology, capillaries are integral to microfluidic devices, which are used for lab-on-a-chip applications, enabling rapid and efficient analysis of small fluid samples. In the energy sector, they are essential in the design of heat pipes used for efficient cooling of electric motors, as they facilitate effective heat transfer and distribution \cite{faghri}. 
Gaining insights into the transport dynamics at the liquid-vapor interface in capillaries within the porous wick structure of heat pipes, as well as their wetting characteristics, is crucial. This understanding plays a significant role in precisely determining the operational conditions of heat pipes and in evaluating the capillary dry-out limitations \cite{rice}. These tiny structures also find applications in ink jet printing technology, where capillary action is used to precisely control the flow of ink onto the printing substrate. Overall, the understanding and manipulation of capillaries and their action are vital in advancing these technologies and improving their efficiency and effectiveness.

In the capillary simulations, accurately modeling the movement of the interface between liquid and vapor in two-phase flow is essential. Various approaches have been developed to achieve this. In computational fluid dynamics, the two primary models for representing the interface are the diffusive and sharp interface models. Diffusive interface models \cite{JACQMIN_2000} represent the interface as a smooth, transitional zone where fluid properties gradually change. In contrast, sharp interface models \cite{Sibley_Nold_Kalliadasis_2013} treat the interface as an extremely thin and well-defined, where fluid properties change abruptly. Each of this methods has its own set of advantages and challenges. Diffusive interface models able to handle complex interface shapes and more numerically stable, but it may sacrifice some physical accuracy and require more computational resources. On the contrary, sharp interface models offer a high level of physical accuracy but can be computationally demanding. 

Furthermore, these models are operationally executed through either interface tracking  or interface capturing methods \cite{Tezduyar}. Interface tracking methods explicitly track the interface's position and shape, using a network of discrete points or a mesh that moves with the fluid. On the other hand, capturing methods implicitly define the interface within the computational grid without explicitly tracking its position. These methods use a field variable (like volume fraction or a level set function) to indicate the presence of different fluids in each computational cell. Examples include the front tracking method for interface tracking, and the volume of fluid and level-set methods for interface capturing.

The selection of a suitable method is dictated by the specific requirements of the simulation, including factors like desired accuracy, available computational resources, and the type of physical phenomena being modeled. In our research, we utilize the level-set method for two-phase flow simulations. This method characterizes the boundary between two fluids using a signed distance function, which has a value of zero at the boundary and varies in the adjacent fluid phases \cite{Osher}. For mathematical discretization, we employ the Extended Discontinuous Galerkin (XDG) method, developed by Kummer \cite{kummer}, which draws on ideas from the extended finite element method (XFEM) \cite{Moës} and also known as cut-cell DG or unfitted DG \cite{bastian}. XDG is notable for its high-order approximation of the boundary and a quadrature technique tailored for domains defined indirectly. The discretization process in this approach is executed using the Symmetric Interior Penalty (SIP) method.

In our simulations, we encounter a significant challenge due to the existence of a three-phase contact line, which presents a complex issue. One of the primary challenges associated with the contact line is the stress singularity that arises when applying classical hydrodynamic models, particularly the no-slip boundary condition. This condition leads to a non-integrable stress singularity at the contact line, resulting in unphysical predictions regarding the forces required to move the contact line \cite{shikhmurzaev}, \cite{HolmgrenKreiss2019}. The classical models fail to account for the complex interactions at the three-phase contact line, which can lead to significant discrepancies between theoretical predictions and experimental observations \cite{shikhmurzaev}, \cite{Qian2006}. 

One of the key aspects of the moving contact line problem is the dependence of the dynamic contact angle on the speed of the contact line. In their work, Eggers et. al. \cite{Eggers2004}  describe how the dynamic contact angle can be expressed in terms of characteristic lengths influenced by the capillary number, which reflects the balance between viscous and surface tension forces. This relationship is critical for understanding how the contact angle evolves as the contact line moves, especially under varying flow conditions. Additionally, authors \cite{Golestanian2002}  discuss the relaxation dynamics of moving contact lines, noting that the behavior of advancing and receding contact lines can differ significantly, which has implications for the modeling of dynamic wetting processes.

Interactions at the interface involve a complex array of physical phenomena that significantly influence the behavior of the contact line. These include surface tension, which manifests as capillary effects, and viscous forces, both of which are critical in determining the flow dynamics near the contact line. Additionally, when volatile liquids are involved, phase changes such as evaporation or condensation become significant, further complicating the modeling process. The cumulative impact of these forces varies depending on the specific properties of the interacting phases, making the accurate modeling of flow behavior near the contact line particularly challenging \cite{Pierre-Gilles}.

Additionally, issues with numerical stability and convergence arise. We must consider the capillary time step limitation \cite{Denner} to ensure the stability of our numerical method. In general, the complex interplay of forces at the contact line, together with the sharp gradients and discontinuities in material properties at the interface, presents a challenge for the convergence of numerical methods. Convergence problems may manifest as numerical solvers failing to find a stable solution, which can halt or significantly slow down a simulation.

To resolve these challenges, we introduce a semi-analytical model designed for modeling the dynamics of a moving contact line with evaporation in a capillary. This model, called a Parameterized Level-Set Method, uses a specifically defined function to approximate the liquid-vapor interface shape. This strategy effectively solves issues associated with singularities and enables a substantial enlargement of the time step size, which significantly decreases computational expenses.
\section{Governing equations}
\label{sec:2}
\subsection{Balance equations in the bulk domain}
\label{subsec:1}
In this study, we examine transient two-phase flow with a moving interface within domain $ \Omega$. This domain is delineated as the time-varying separate partitions of fluid bulk phases 
$\mathfrak{A(t)}$ and $\mathfrak{B(t)}$, along with the moving interface $\mathfrak{I(t)}$:
\begin{equation}
    \Omega = \mathfrak{A(t)}~ \cup~ \mathfrak{I(t)} ~\cup ~ \mathfrak{B(t)}.
\end{equation}
The transient incompressible Navier–Stokes equations, coupled with the heat equation, are employed to describe the problem. Balance equations for continuity, momentum and temperature in the bulk domain  $\Omega \setminus \mathfrak{I}$ are given as:
\begin{subequations}
	\begin{align}
    &\nabla \cdot \vec{u} =0,  \\ 
    &\rho \left( \frac{\partial \vec{u} }{\partial t} + \vec{u} \cdot \nabla \vec{u} \right) = - \nabla p + \mu \nabla \cdot \left( \nabla \vec{u} + \left( \nabla \vec{u} \right)^T \right), \\
    &\rho c  \left(\frac{\partial T} {\partial t}+\vec{u} \cdot \nabla T \right) = k\Delta T.
	\end{align}
\end{subequations}
Here $\rho$, represents the density, which is constant within both phases; $\mu$, $k$ and $c$  are the dynamic viscosity, thermal conductivity, and heat capacity, respectively. The following assumptions are made: the fluid under consideration is Newtonian; the heat flux is isotropic and can be described using Fourier's law $\vec{q}=-k \nabla T$. The unknowns in the given equations are pressure $p=p(\vec{x},t)$, velocity $ \vec{u}=\vec{u}(\vec{x},t)$ and temperature $T=T(\vec{x},t)$. 

Considered boundary conditions are Dirichlet and Neumann: 
\begin{subequations}
    \begin{align}      
     &\vec{u} = \vec{u}_D ~~\textrm{and}~~ T=T_D~~\textrm{on}~~ \partial \Omega_D,\\
    & \ -p \vec{n}_{\partial \Omega} + \mu \left( \nabla \vec{u} + \left( \nabla \vec{u} \right)^T \right)  \vec{n}_{\partial \Omega} = -p_{ext}\vec{n}_{\partial \Omega} ~~\textrm{and}\\
    &  -k \nabla T \cdot \vec{n}_{\partial \Omega} = \vec{q}_N  \vec{n}_{\partial \Omega} ~~\textrm{on}~~ \partial \Omega_N.
    \end{align}
\end{subequations}
Here, $ \vec{u}_D$, $T_D$, $\vec{q}_N$ and $p_{ext}$ are given boundary values. On the boundary $\partial \Omega$ the normal $\vec{n}_{\partial \Omega}$ is oriented outward.
The initial value problem is resolved by defining initial conditions for velocity and temperature:
\begin{subequations}
    \begin{align}      
     &\vec{u}(\vec{x},0)=\vec{u}_0(\vec{x}) ~~~~\textrm{for}~~\vec{x} \in \Omega \setminus \mathfrak{I},\\
     & T(\vec{x},0)=T_0(\vec{x}) ~~~~\textrm{for}~~\vec{x} \in \Omega \setminus \mathfrak{I}.
    \end{align}
\end{subequations}

\subsection{Balance equations on the interface}
\label{subsec:2}
For the mass balance at the interface we assume that the interface itself does not have any additional mass and  has velocity $\vec{s}$:
\begin{equation}
    \llbracket \rho \left(\vec{u}  - \vec{s} \right) \cdot \vec{n_{ \mathfrak{I}}} \rrbracket = 0. 
\end{equation}
Considering this equation in $\mathfrak{A}$ and $\mathfrak{B}$ phases, it can be rewritten in the following form
\begin{equation}
    \rho_{\mathfrak{A}} \left(\left(\vec{u}_{\mathfrak{A},\mathfrak{I}} - \vec{s}  \right)\cdot \vec{n_{\mathfrak{I}}}  \right) = \rho_{\mathfrak{B}} \left(\left(\vec{u}_{\mathfrak{B},\mathfrak{I}} - \vec{s}  \right)\cdot \vec{n_{\mathfrak{I}}}  \right) =  \dot m.
    \label{mass_flux_interface}
\end{equation}
Here, $\dot m$ is the mass transfer rate. 
By removing the surface velocity $\vec{s}$ from the equation, the mass balance expressed in terms of the mass transfer rate is as follows:
\begin{equation}
    \llbracket \vec{u}  \rrbracket \cdot \vec{n_{ \mathfrak{I}}} = -\dot m  \llbracket \rho^{-1} \rrbracket.
    \label{conti_interface}
\end{equation}
Given the assumption of a no-slip condition at the interface, the momentum balance equation for the interface is presented as
\begin{equation}
    -\dot m \llbracket \vec{u}\rrbracket + \llbracket -p I + \mu \left(\nabla \vec{u} + \left(\nabla \vec{u} \right)^T \right) \rrbracket \vec{n_{ \mathfrak{I}}} = - \sigma \kappa \vec{n_{ \mathfrak{I}}}.
\end{equation}
Here, $\sigma$ and $\kappa$ are surface tension and mean interface curvature, respectively.
Considering the heat equation, we assume the temperature at the interface to be continuous and equivalent to the saturation temperature $T_{sat}$: 
\begin{subequations}
    \begin{align}
    & \llbracket T \rrbracket = 0, \\
    & T_{ \mathfrak{I}} = T_{sat}.
    \end{align}
\end{subequations}
The heat balance equation at the interface is expressed as
\begin{equation}
    \llbracket -k \nabla T \rrbracket \cdot \vec{n}_{ \mathfrak{I}} = \dot m h_{vap}.
\end{equation}
Here, $h_{vap}$ is an enthalpy of evaporation.

\subsection{Generalized Navier Boundary Condition}
\label{subsec:3}
\begin{figure}[h]
\centering
\includegraphics[scale=0.15]{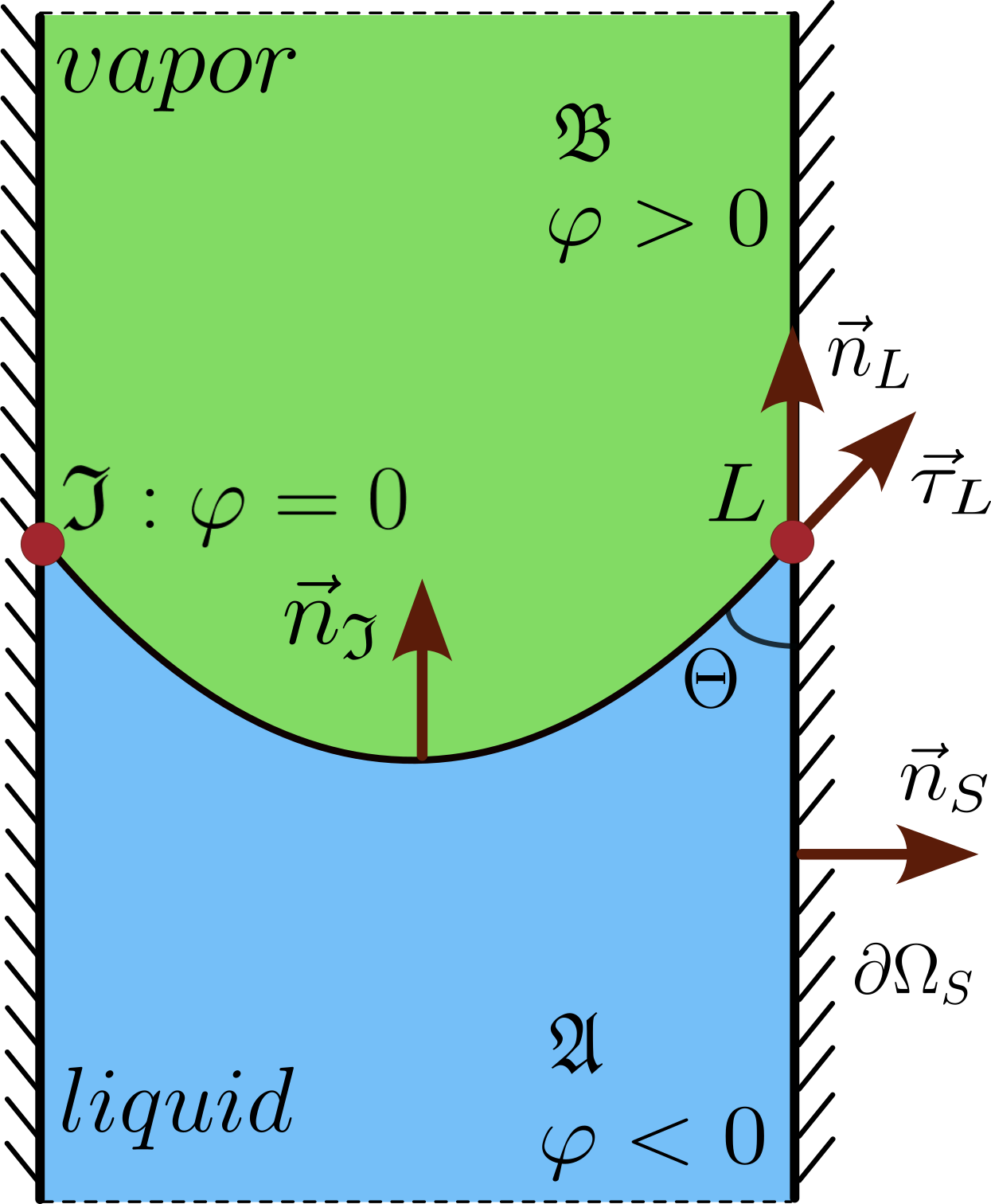}
\caption{Contact line in the capillary rise problem}
\label{fig:contact_line} 
\end{figure}

On the wall boundaries, the Navier-slip boundary condition is implemented. This condition fundamentally introduces suitable dissipative effective forces at both the slip wall, denoted as $\partial \Omega_S$, and the contact line $L$ (Figure \ref{fig:contact_line}). On this figures $\Theta$ is contact angle between the interface and the wall; $\vec{n}_S$ is the normal to a slip wall $\partial \Omega_S$; $\vec{n}_L$ is the normal to $L$ and tangential to the wall $\partial \Omega_S$; $\vec{\tau}_L$ is the normal to $L$ and tangential to $\mathfrak{I}$; $\vec{n}_{\mathfrak{I}}$ is the normal to $\mathfrak{I}$.

In this case, at $\partial \Omega_S$ instead of the no-slip condition, the slip condition is imposed alongside the no-penetration condition. Subsequently, we have 
\begin{subequations}
    \begin{align}
    &\mu \textbf{P}_S  \left( \nabla \vec{u} + \left( \nabla \vec{u} \right)^T \right) \vec{n}_{s} = \vec{f}_S ~~~~\textrm{on} ~~~~\partial \Omega_S,\\
    &\vec{u} \cdot \vec{n}_S = 0~~~~\textrm{on} ~~~~\partial \Omega_S. 
\end{align}
\label{navier_slip}
\end{subequations}
Here, $\textbf{P}_S:=I - \vec{n}_{\partial{\Omega}} \otimes \vec{n}_{\partial{\Omega}}$ denotes the orthogonal projection onto the wall $\partial \Omega_S$.

\noindent
A dissipative friction force $\vec{f}_S$ is defined as
\begin{equation}
    \vec{f}_S = -\beta_S \textbf{P}_S \vec{u}.
\end{equation}
Here, $\beta_S$ is the phase friction coefficient, which introduces a slip length $l_s$, where $\beta_s=\mu / l_s$. When $\beta_S=0$, a free-slip boundary condition is applied, while as $\beta_S \to\infty$, the condition transitions to no-slip.

An additional dissipative force is introduced at the moving contact line $L(t)$: 
\begin{equation}
    \vec{f}_{L,diss} = -\beta_L U_L \vec{n}_L,
    \label{dissipative_force}
\end{equation}
where coefficient $\beta_L \geq 0$ and $U_L=\vec{u}\cdot \vec{n}_L $ is a contact line velocity.

Combining equation (\ref{dissipative_force}) with Young's equation \cite{smuda}, one obtains equation for the resulting effective force acting at the contact line
\begin{equation}
    \vec{f}_{L} = -\beta_L U_L \vec{n}_L + \sigma \cos{\theta_{\mathrm{stat}}},
\end{equation}
where $\theta_{\mathrm{stat}}$ is the static contact angle.

\noindent
From momentum conservation, the force balance condition must hold at $L$ 
\begin{equation}
    \textbf{P}_S \textbf{S}_{\mathfrak{I}} \tau_L = \vec{f}_L,
    \label{balance_eq}
\end{equation}
where $\textbf{S}_{\mathfrak{I}}=\sigma \textbf{P}_{\mathfrak{I}}$ is interface stress tensor given as the isotropic part. By rearranging the left-hand side of equation (\ref{balance_eq}), one obtains  
\begin{equation}
    \sigma (\cos{\theta_{\mathrm{stat}}}-\cos{\theta}) = \beta_L U_L~~~~\textrm{on} ~~~~L,
    \label{force_balance}
\end{equation}
where $\theta$ is the current contact angle.

For $\beta_L=0$, combining equations (\ref{force_balance}) and (\ref{navier_slip}) leads to the Generalized Navier boundary condition.
\section{Numerical method}
\label{sec:3}
\subsection{Discontinuous Galerkin Method(DG)}
\label{subsec:1}
In our discussion on the DG method, we start with foundational definitions relevant to its application. The computational domain $\Omega \in R^D $ is a polygonal and simply connected. We construct a numerical mesh $\mathfrak{K}_h = {K_1, \ldots, K_N}$ (Figure \ref{fig:DG_Grid}), that comprehensively overlays the domain $\overline{\Omega}$ through a series of non-overlapping cells, i.e., $\overline{\Omega} = \bigcup_j \overline{K}j$ and $\int{K_j \cap K_l} 1dV = 0$ for $l \neq j$. The parameter $h$ represents the maximum diameter among the cells.

In DG,  some field $\psi$ is approximated by $ \psi_h(\vec{x}, t)$ for each numerical cell $K_j$ as follows 
\begin{equation}
    \psi_j(\vec{x},t) =\sum_{n=1}^{N_k} \tilde{\psi}_{j,n}(t)\phi_{j,n}(\vec{x}).
\end{equation}
Here $\tilde{\psi}_{j}=( \tilde{\psi}_{j,n})_{n=1,...,N_k}$ is unknown degrees of freedom (DOF) of the local solution  and $\phi_{j}=( \phi_{j,n})_{n=1,...,N_k} \in \mathbb{P}_k(\{K_j\}) $ is a cell-local polynomial basis.

\subsection{Extended Discontinuous Galerkin Method(XDG)}
\label{subsec:2}
The DG method is adapted into the Extended Discontinuous Galerkin Method (XDG) to effectively tackle the complexities of two-phase flow problems \cite{kummer}.  In this case the existence of a dividing interface $\mathfrak{I}$ with $\Omega = \mathfrak{A(t)} \cup \mathfrak{I(t)} \cup \mathfrak{B(t)}$ is allowed.  With the incorporation of the interface $\mathfrak{I}$, it becomes possible to designate specific cells to correspond exclusively with a particular phase $\mathfrak{s(t)} \in \{\mathfrak{A(t)}, \mathfrak{B(t)}\}$. However, in regions occupied by the interface, both phases coexist within the same cells. To accurately represent this phenomenon within the computational mesh, cut-cells and a corresponding cut-cell mesh are introduced.

\begin{figure}[h]
  \centering
  \begin{subfigure}[b]{0.46\textwidth}
    \centering
    \includegraphics[width=\textwidth]{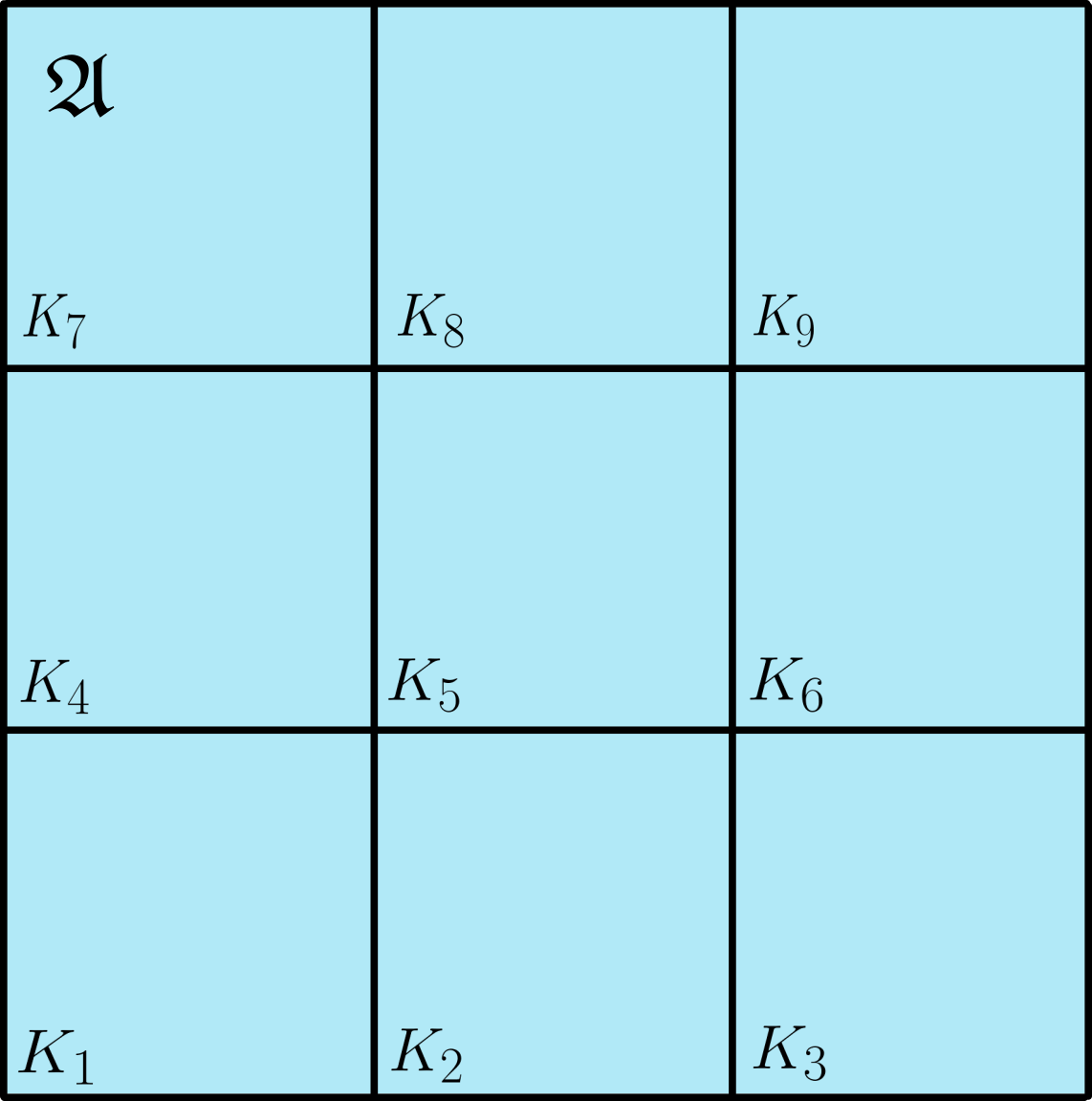}
    \caption{Standard background mesh}
    \label{fig:DG_Grid}
  \end{subfigure}
  \hfill
  \begin{subfigure}[b]{0.46\textwidth}
    \centering
    \includegraphics[width=\textwidth]{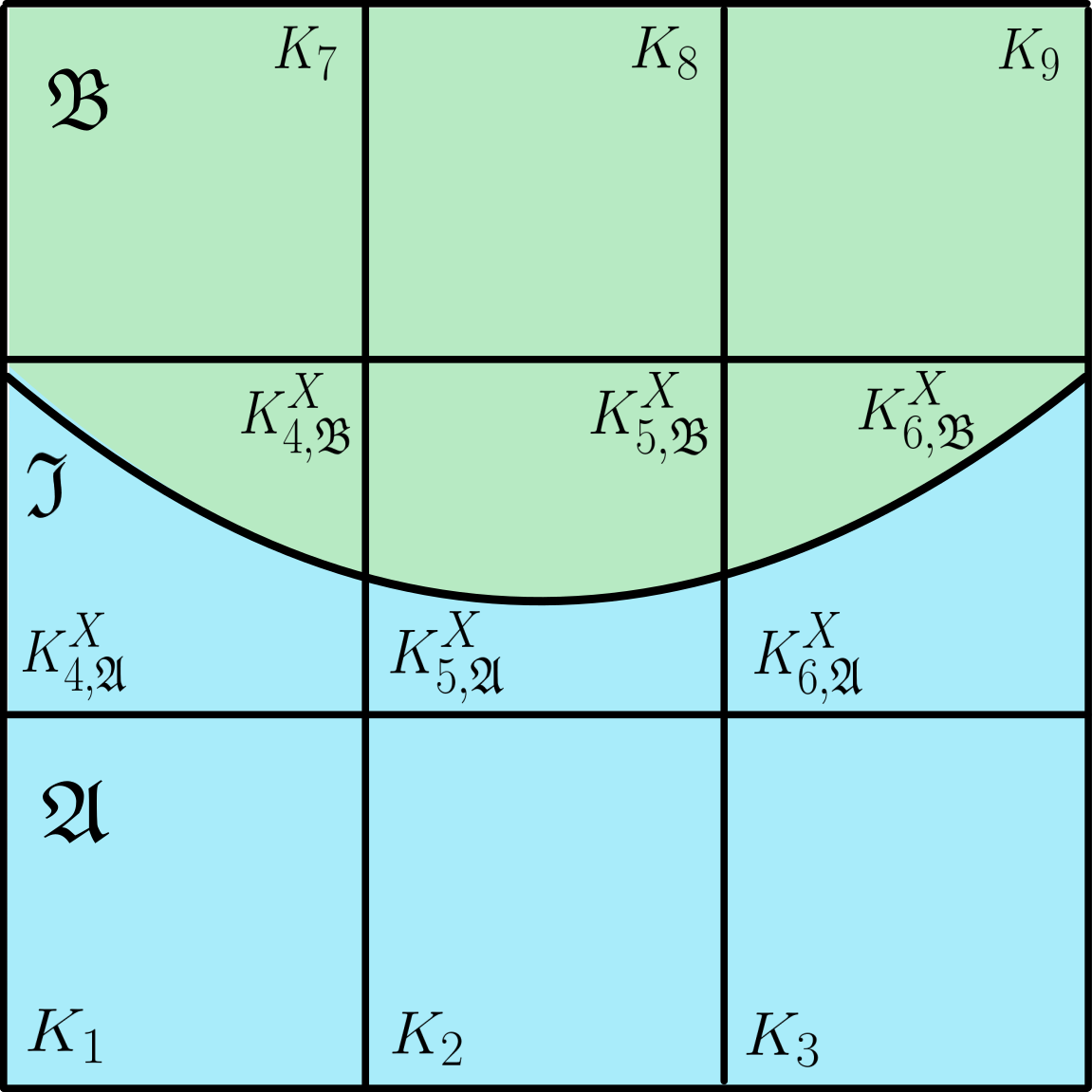}
    \caption{Cut-cell mesh}
    \label{fig:XDG_Grid}
  \end{subfigure}
  \caption{Types of meshes}
  \label{fig:droplet_wall}
\end{figure}

The concept of time-dependent cut-cells is encapsulated by the notation $K_{j,s}^X(t):=K_j \cap \mathfrak{s(t)}$. The set of all cut-cells $K_{j,s}^X(t)$ composes  the time-dependent cut-cell mesh $ \mathfrak{k_h^X(t)} = \{K_{1,\mathfrak{A}}^X(t), K_{1,\mathfrak{B}}^X(t),...,K_{J,\mathfrak{A}}^X(t), K_{J,\mathfrak{B}}^X(t)\} $ (Figure \ref{fig:XDG_Grid}).

In cells containing a single phase, the standard DG approximation is employed. In contrast, when a cell encompasses two phases, the XDG approximation is applied. Within a single cell containing two phases, the local representation of a field property  $\psi_j$  is described by the following:
\begin{equation}
    \psi_j(\vec{x},t) =\sum_{n=1}^{k}  \underbrace{\tilde{\psi}_{j,n, \mathfrak{A}}(t) \phi_{j,n}(\vec{x})1_{\mathfrak{A}}(\vec{x}, t)}_{\textrm{phase} ~\mathfrak{A}} ~ + ~\underbrace{\tilde{\psi}_{j,n, \mathfrak{B}}(t) \phi_{j,n}(\vec{x})1_{\mathfrak{B}}(\vec{x}, t)}_{\textrm{phase} ~\mathfrak{B}}, ~~x \in K_j,
\end{equation}
where the characteristic functions $1_{\mathfrak{A}}(\vec{x}, t)$ and $1_{\mathfrak{B}}(\vec{x}, t)$ are defined as 
\begin{equation}
    1_{\mathfrak{A}} :=
    \begin{cases}
        1 &\text{in $\mathfrak{A}\left(t\right)$}\\
        0 &\text{in $\mathfrak{B}\left(t\right)$}
    \end{cases},
\end{equation}

\begin{equation}
    1_{\mathfrak{B}} :=
    \begin{cases}
        0 &\text{in $\mathfrak{A}\left(t\right)$}\\
        1 &\text{in $\mathfrak{B}\left(t\right)$}
    \end{cases}.
\end{equation}
The cut-polynomial basis for a phase $\mathfrak{s}(t) \in {\mathfrak{A}(t), \mathfrak{B}(t)}$ is given by $ \psi^X_{j,n, \mathfrak{s}}(\vec{x}, t) =  \phi_{j,n}(\vec{x}) 1_{\mathfrak{s}}( \vec{x}, t) $ 
 and corresponding coefficients are denoted by $\psi_{j,n,\mathfrak{s}}(t)$.

For the spatial discretization of a two-phase problem, we refer the reader to the work Rieckmann et. al \cite{Rieckmann}.

\subsection{Parameterized Level-Set representation}
\label{subsec:4}
\subsubsection{Level-Set Method}
\label{subsubsec:1}
We describe the moving interface using the level-set function $\varphi(\vec{x},t)$ (Figure \ref{fig:contact_line}).
For interface representation an implicit form is used 
\begin{equation}
 \varphi(\vec{x}, t) = 0, ~~\vec{x}  \in \mathfrak{I}(t).
\end{equation}
The level-set evolution equation is obtained through temporal differentiation of the preceding equation:
\begin{equation}
\frac{\partial \varphi}{\partial t} + \vec{s} \cdot \nabla \varphi = 0.
\label{ls_evol}
\end{equation}
Geometric interface properties including normal $\vec{n}_{\mathfrak{I}}$ and mean curvature $\kappa$ are computed as
\begin{subequations}
    \begin{align}
     & \vec{n}_\mathfrak{I} = \frac{\nabla \varphi}{|\nabla \varphi|}, \\
   & \kappa = \nabla \cdot \left( \frac{\nabla \varphi}{|\nabla \varphi|} \right).
    \end{align}
\end{subequations}

\subsubsection{The concept of Parameterized Level-Set Method}
\label{subsubsec:2}
The primary contribution of this work is the Parameterized Level-Set Evolution Method, which addresses the challenges posed by singularities at the three-phase contact line and offers the ability to overcome the capillary time-step restriction while reducing computational costs.

This method is predicated on the assumption that the interface adopts a parameterized shape and retains this parametric configuration throughout each time step. Thus, the level-set field can be described by a \emph{global} Ansatz (given below), which depends only on three time-dependent parameters $a(t), b(t), c(t)$. 
We consider two different Ansatz functions for the interface approximation:
\begin{enumerate}
\item{Elliptic Ansatz: In the first case, which involves an elliptic function, the corresponding level-set function is given by the following expression:
  \begin{equation}
    \varphi_{\textrm{ell}}(a,b,c,x,y) = y - c +  \sqrt{b^2  \cdot  \left( 1 - \frac{x^2}{a^2} \right)}~.
    \label{level-set func}
\end{equation}
Here, $a(t)$, $b(t)$ represent the semi-major and semi-minor axes, respectively, and $(0, c(t))$ denotes the center point of the ellipse.}
\item{Polynomial Ansatz: For the second case with fourth-order polynomial, the level-set function is defined as follows:
\begin{equation}
    \varphi_{\textrm{poly}}(a,b,c,x,y) = y - a \cdot x^4 - b \cdot x^2 - c .
    \label{level-set func polyn4}
\end{equation}}
\end{enumerate}

\noindent
We then employ one of these global Ansatz functions to represent the level-set function, i.e.,
\begin{equation}
    \varphi(t,x,y)= \varphi_*(a(t),b(t),c(t),x,y),
\end{equation}
where $()_*$ stands either for $"\textrm{ell}"$ or $"\textrm{poly}"$ Ansatz. The time-dependent parameters $a(t),b(t),c(t)$ are the actual degrees-of-freedom of the Parameterized Level-Set Method, which will be outlined below.

In our formulations, we assume that only the parameters $a(t)$, $b(t)$ and $c(t)$ are time-dependent and make the approximation that these parameters have a linear dependence on time:
\begin{equation}     
     a(t) = a_0 + k_a  t,~~b(t) = b_0 + k_b  t,~~c(t) = c_0 + k_{c}  t.
    \label{linear_depend} 
\end{equation}
Here, $k_a,  k_b$ and $k_{c}$ are the slopes of the functions $a(t), ~b(t)$ and $c(t)$, respectively.

We discretize the time derivative in equation (\ref{ls_evol}) using an Explicit Euler Method \cite{num_recipes}, yielding:

\begin{equation} 
    \frac{\varphi_1 - \varphi_0}{\Delta t} + \vec{s} \cdot \nabla \varphi_0 = 0, \label{time-derivative} 
\end{equation}
 where $\varphi_1$ is the level-set function at time $t_1=t_0+\Delta t$, parameterized by $(a_0 + k_a \Delta t,~b_0+ k_b \Delta t,~c_0 + k_c \Delta t)$, and $\varphi_0$ is the level-set function at time $t_0$.

Next, we select the function Ansatz from one of the given options (\ref{level-set func}) or (\ref{level-set func polyn4}) and substitute it into equation (\ref{time-derivative}). This introduces three unknown parameters $(k_a, k_b, k_c)$, which we determine by solving the following minimization problem:

\begin{equation}
F(k_a, k_b, k_c)=\oint_{\mathfrak{I}} \left( \frac{\varphi_1 - \varphi_0}{\Delta t} + \vec{s}\cdot \nabla \varphi_0 \right)^2dS \rightarrow min,
 \label{F_min}
\end{equation}
where the integral is evaluated along the interface $\mathfrak{I}$.

A critical component of our study involves selecting a minimization method to optimize parameters at each time step. 

\subsubsection{Adaptive Moment Estimation Method (Adam)}
\label{subsubsec:3}
To solve the minimization problem (\ref{F_min}), we chose Adam's method \cite{kingma2017adam}, which allowed us to achieve the quickest convergence. This method, originally designed to efficiently solve large-scale optimization problems, adjusts the step size of each parameter based on its gradient history, thereby improving overall optimization efficiency. As a result, it converges more quickly and reliably than other minimization methods, such as Stochastic Gradient Descent and Momentum Gradient Descent Methods \cite{Ruder}, making it more efficient for the current minimization problem.

We define $\vec{w}_i=[k_{a},k_{b},k_{c}]_i$ as the vector of parameter values at iteration $i$. Consequently, the gradient of our loss function $F(k_{a_i},k_{b_i},k_{c_i})$ at iteration $i$ is denoted by $\frac{\partial F(\vec{w}_i)}{\partial \vec{w}_i} $.
At each iteration of the Adam algorithm, we compute the gradient $ g_i= \frac{\partial F(\vec{w}_{i-1})}{\partial \vec{w}_{i-1}}  $.
We then update exponential moving averages of the gradient, denoted by ${m_i}^{(d)}$ (the first moment estimate), and of its square, denoted by ${v_i}^{(d)}$ (the second moment estimate). The hyperparameters $\beta_1$ and $\beta_2$ control the decay rates of these averages.
\begin{subequations}
    \begin{align}  
    & {m_i}^{(d)}=\beta_1 \cdot {m_{i-1}}^{(d)} + (1-\beta_1){g_i} ^{(d)}, \\
     & {v_i}^{(d)}=\beta_2 \cdot {v_{i-1}}^{(d)} + (1-\beta_2)({g_i} ^{(d)})^2~~~~\textrm{for} ~~d=1,2,3.
    \end{align}
\end{subequations}
Here, ${m_i}^{(d)}$, ${v_i}^{(d)}$ and ${g_i} ^{(d)}$ denote the 
$d$-th components of the first, second moment vectors and gradient of the loss function, respectively.

In the initial phase of optimization, limited data can cause these moment estimates to be biased toward zero, which may lead to large step sizes and unstable updates. To mitigate this issue, we apply the following bias-correction factors: 
\begin{subequations}
    \begin{align} 
   & \hat{m_i}^{(d)}= \frac{m_i^{(d)}}{1-{\beta_1}^i}, \\
   & \hat{v_i}^{(d)}= \frac{v_i^{(d)}}{1-{\beta_2}^i} ~~~~\textrm{for} ~~d=1,2,3.
    \end{align}
\end{subequations}
Finally, the parameter update rule is:
 \begin{equation}
     \vec{w}_{i}^{(d)} = \vec{w}_{i-1}^{(d)}- \lambda \frac{\hat{m_i}^{(d)}} {\sqrt{\hat{v_i}^{(d)}} + \epsilon } ~~~~\textrm{for} ~~d=1,2,3.
 \end{equation}

\noindent
The specific parameters used for this method are summarized in the Table \ref{tab:adam_values}, where $\lambda$ - learning rate, $\epsilon$  - term added to the denominator to improve numerical stability, $\beta_1$ and $\beta_2$ - coefficients used for computing running averages of gradient and its square respectively.

\begin{table}[h]
\caption{Parameter values for Adam method}
\label{tab:adam_values}     
\begin{tabular}{p{2.2cm}p{2.2cm}p{2.2cm}p{2.2cm}p{2.2cm}}
\hline\noalign{\smallskip}
Parameter & $\beta_1$ & $\beta_2$ & $\lambda$ & $\epsilon$   \\
\noalign{\smallskip}\hline\noalign{\smallskip}
  & 0.99 & 0.999 &  0.01 & $10^{-6}$\\

\noalign{\smallskip}\hline\noalign{\smallskip}
\end{tabular}
\end{table}

\section{Numerical results}
\label{sec:4}

\subsection{Capillary rise}
\label{subsec:1}

\subsubsection{Setting of the problem}
\label{subsubsec:1}
\begin{figure}[h]
\centering
\includegraphics[scale=.12]{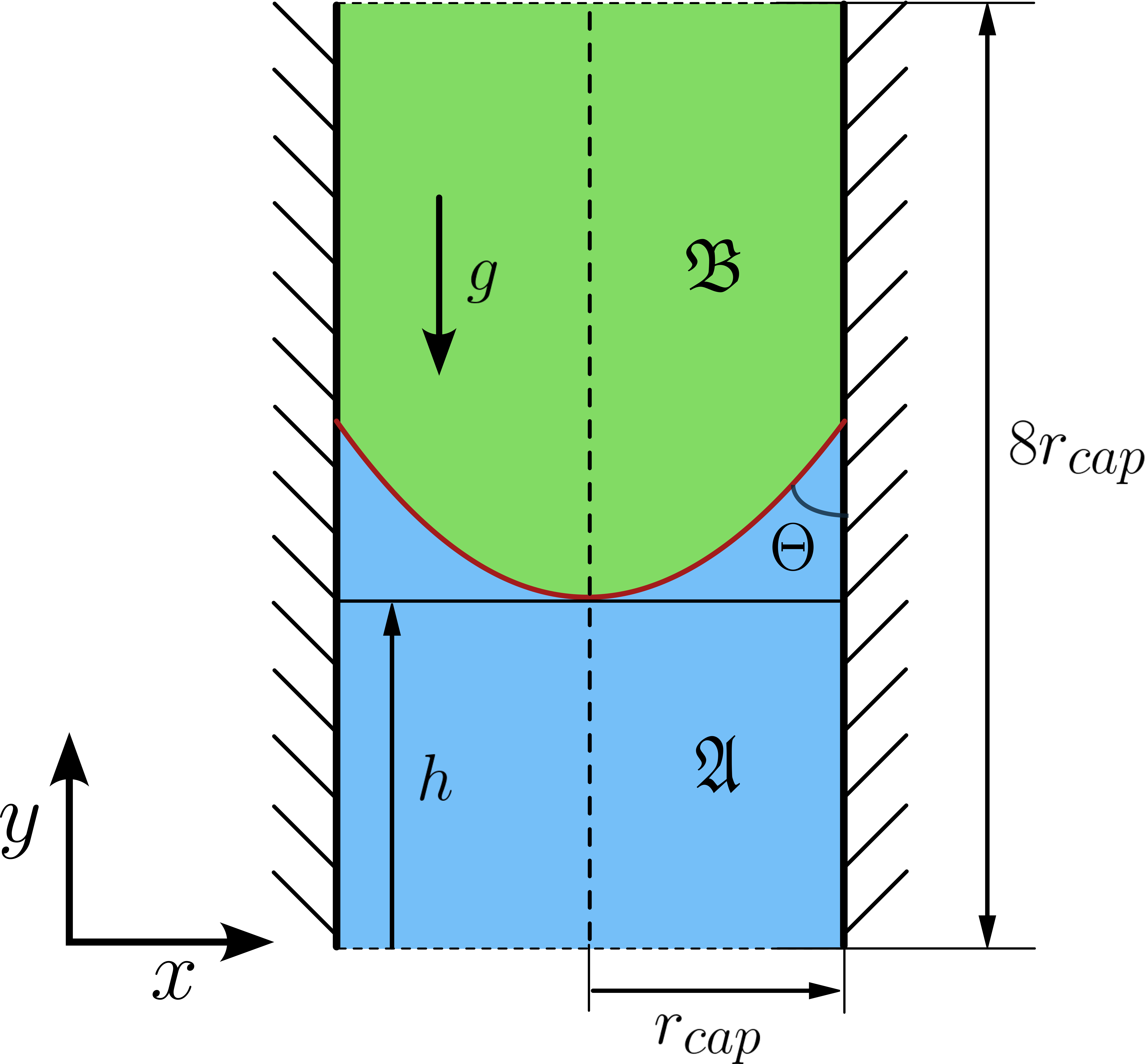}
\caption{Computational domain for a capillary rise calculation}
\label{fig:capillary_rise} 
\end{figure}

The setting of the problem is analogous to that presented in the work of Gründing et. al. \cite{smuda_capillary}. Computational domain is presented on the Figure \ref{fig:capillary_rise}. In the same manner, we choose physical parameters such that the non-dimensional parameter $\Omega_\textrm{iner}$ (see Table \ref{tab:capillary_rise})
\begin{equation} 
\Omega_\textrm{iner} = \sqrt{\frac{9 \sigma \cos \theta \mu^2}{\rho^3 g^2 R^5}} = 1,
\end{equation}
 where $\sigma$ represents the surface tension, $\theta$ the static contact angle, $\mu$ the fluid viscosity, $\rho$ the fluid density, $g$ the magnitude of gravity, and $R$ the inner tube radius. The parameter $\Omega_\textrm{iner}$ evaluates the influence of inertia; when it decreases, oscillations increase.
 
To enable a direct comparison with the work of Gründing et. al. \cite{smuda_capillary}, we set the contact angle  $\theta_e$ to $30^\circ$.

\begin{table}[h]
\caption{Physical parameters for capillary rise calculation}
\label{tab:capillary_rise}     
\begin{tabular}{p{1cm}p{1.6cm}p{1.7cm}p{1.7cm}p{1.6cm}p{1.7cm}p{1.7cm}}
\hline\noalign{\smallskip}
$\Omega_\textrm{iner}$ & $r_\mathrm{cap}$, $\mathrm{m}$ & $ \rho$, $\mathrm{kg/m^3}$ & $ \mu$, $\mathrm{Pa \cdot s}$ & $g$, $\mathrm{m/s^2}$ & $\sigma$, $\mathrm{N/m}$ & $\theta_e$, $^0$  \\
\noalign{\smallskip}\hline\noalign{\smallskip}
1 & 0.005 & 83.1 & 0.01 & 4.17 & 0.04 & 30\\

\noalign{\smallskip}\hline\noalign{\smallskip}
\end{tabular}
\end{table}

We can calculate the stationary capillary rise height as
\begin{equation}
    h_{\infty}^{\mathrm{apex}} = h_{\mathrm{Jurin,2D}} - \hat{h},
\end{equation}
where
\begin{equation} 
h_\textrm{Jurin, 2D} = \frac{\sigma \cos \theta}{R \rho g},
\end{equation} 
and 
\begin{equation} 
\hat{h} = \frac{R }{2 \cos \theta} \left( 2 - \sin \theta - \frac{\arcsin(\cos \theta)}{\cos \theta}\right ) .
\end{equation} 
Here, the height correction $\hat{h}$ accounts for the liquid volume in the interface region, i.e., the area between the solid apex height and the solid interface line.

We consider a symmetric configuration where a free-slip boundary condition with contact angle $\theta_e=90^\circ$ is imposed along the symmetry plane. The inflow and outflow boundary conditions are imposed at the two ends of the capillary, with $\mu \left( \partial_n \mathbf{u} + \nabla \mathbf{u}_n \right) - p \mathbf{n}_{\partial D} = 0$. Navier-slip boundary condition is imposed on the wall with a slip length $L=R/5$. For this simulation we use a polynomial degree $k = 2$  for velocity and $k - 1$ for pressure. The computational grid consists of equidistant cells defined by the grid size  $h$. Similar to the work of Gründing et. al. \cite{smuda_capillary}, we perform an initial calculation to determine the starting position of the interface.

\subsubsection{Results}
\label{subsubsec:2}
\begin{figure}[h]
  \centering
  \begin{subfigure}[b]{0.48\textwidth}
    \centering
    \includegraphics[width=\textwidth]{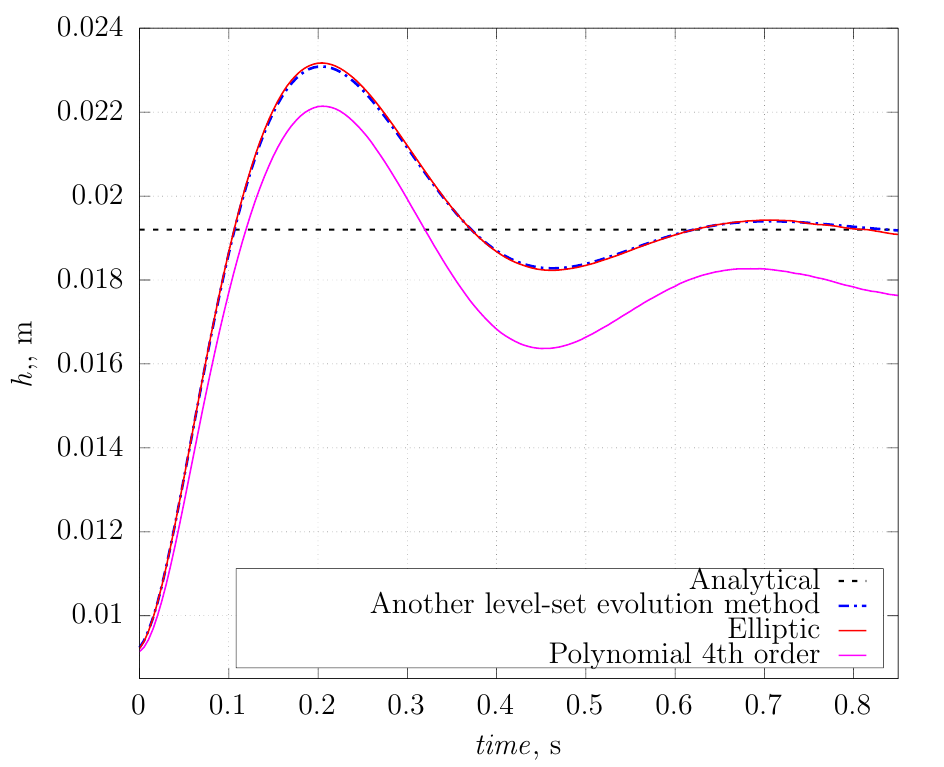}
    \caption{Capillary height}
    \label{fig:capillary_height}
  \end{subfigure}
  \hfill
  \begin{subfigure}[b]{0.48\textwidth}
    \centering
    \includegraphics[width=\textwidth]{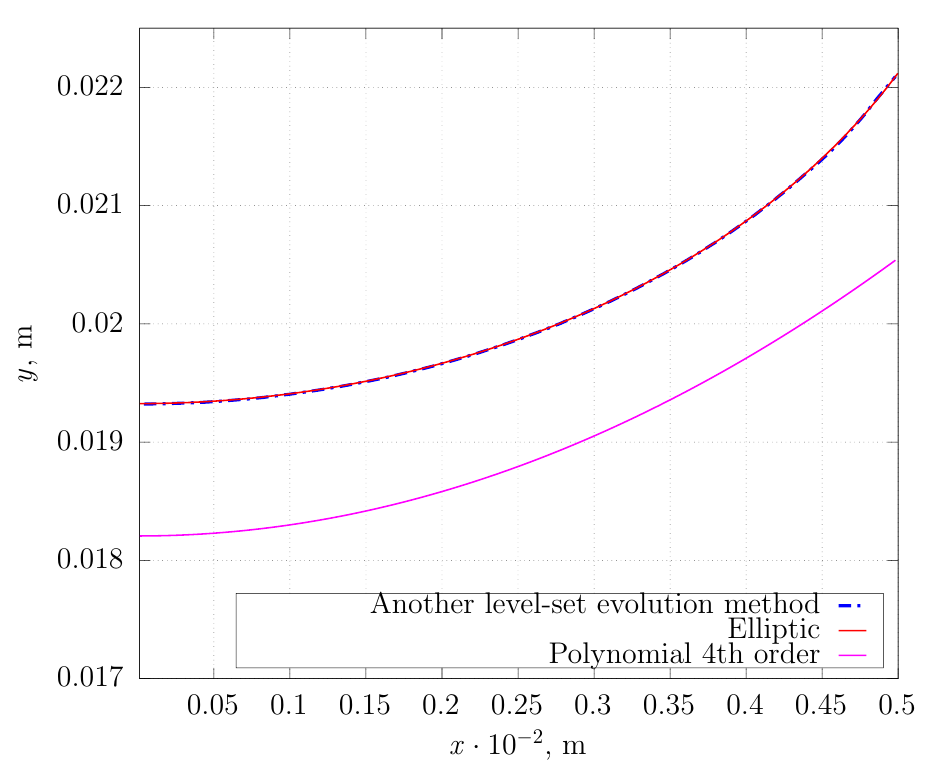}
    \caption{Interface shape}
    \label{fig:capillary_interface}
  \end{subfigure}
  \caption{Capillary rise calculation using different methods for level-set field representation}
  \label{fig:capillary}
\end{figure}

We compare the results obtained with those presented in the work Gründing et. al. \cite{smuda_capillary}. These results are obtained on the finest grid with a size $R/32$. It can be seen that the use of the elliptic Ansatz allows us to obtain a capillary rise height similar to the results reported in the literature, which agrees well with the theoretical value (Figure \ref{fig:capillary_height}). The use of polynomial Ansatz leads to a static value of capillary rise height, which is different from the theoretical value. Furthermore, a deviation in the equilibrium contact angle was observed when using polynomial Ansatz $\varphi_\textrm{poly}$, compared to the prescribed value of 30 degrees. In contrast, the elliptic Ansatz $\varphi_\textrm{ell}$ accurately reproduces the prescribed contact angle (Figure \ref{fig:capillary_interface}). 

\subsection{Capillary rise with evaporation}
\label{subsec:2}

\subsubsection{Setting of the problem}
\label{subsubsec:1}
Compared to the previously considered capillary rise problem, our current study examines a capillary rise problem with evaporation at the interface. Accordingly, we have specified the thermal parameters for the liquid and vapor phases in Table \ref{tab:capillary_rise_evap}. In our calculations we rescale temperature as $\theta = (T-T_\textrm{sat})/\Delta T$.

\begin{table}[h]
\caption{Thermal parameters for capillary rise calculation with evaporation}
\label{tab:capillary_rise_evap}     
\begin{tabular}{p{2.0cm}p{2.0cm}p{2.0cm}p{2.0cm}p{2.0cm}p{1.1cm}p{1.1cm}}
\hline\noalign{\smallskip}
$c_\textrm{l}$ & $c_\textrm{v}$ & $ k_\textrm{l}$ & $ k_\textrm{v}$ & $h_\textrm{vap}$ & $T_\textrm{sat}$  \\
\noalign{\smallskip}\hline\noalign{\smallskip}
 $\mathrm{J / (kg \cdot K)}$ & $\mathrm{J / (kg \cdot K)}$  &  $\mathrm{W / (m \cdot K)}$ &  $\mathrm{W / (m \cdot K)}$ &  $\mathrm{J / kg}$ &  $\mathrm{K}$  \\
\noalign{\smallskip}\hline\noalign{\smallskip}
2000 & 740 & 0.48 & 0.0251 & $10^6$ & 373 \\

\noalign{\smallskip}\hline\noalign{\smallskip}
\end{tabular}
\end{table}
The same way as we did with a previously considered capillary rise setting, we consider a symmetric domain, the boundary conditions remains the same, except a wall where we specify a heat flux value equal to 200 $\mathrm{W / m^2}$. Similar to the work of Gründing et. al. \cite{smuda_capillary}, we perform an initial calculation to determine the starting position of the interface. 

Similar to our previous capillary rise setting, we consider a symmetric domain with consistent boundary conditions, except for the wall where we specify a heat flux value of 200 $\mathrm{W / m^2}$. Following the previous approach, we conduct an initial calculation to determine the starting position of the interface.

\subsubsection{Results}
\label{subsubsec:2}
\begin{figure}[h]
  \centering
  \begin{subfigure}[b]{0.48\textwidth}
    \centering
    \includegraphics[width=\textwidth]{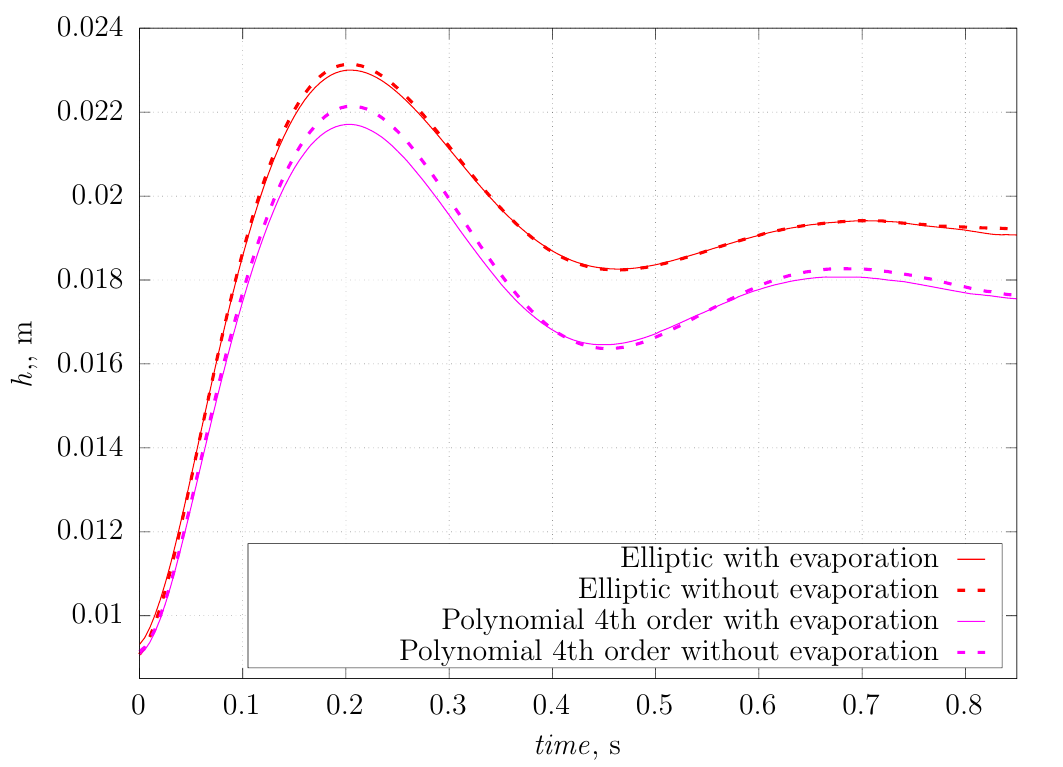}
    \caption{Capillary height}
    \label{fig:capillary_rise_evap_height}
  \end{subfigure}
  \hfill
  \begin{subfigure}[b]{0.48\textwidth}
    \centering
    \includegraphics[width=\textwidth]{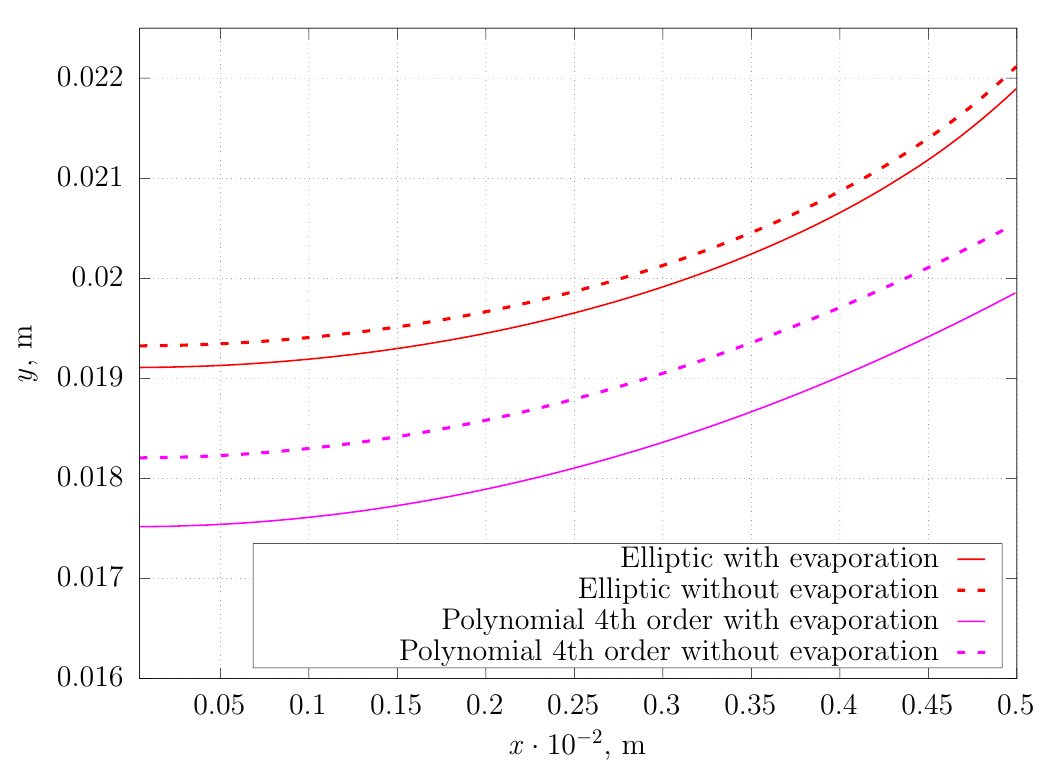}
    \caption{Interface shape}
    \label{fig:capillary_rise_evap_interface}
  \end{subfigure}
  \caption{Capillary rise with evaporation calculation using different methods for level-set field representation}
  \label{fig:capillary_rise_evapy}
\end{figure}

We compared the capillary rise height in a setting with evaporation at the interface to that in a setting without evaporation (Figure \ref{fig:capillary_rise_evap_height}, \ref{fig:capillary_rise_evap_interface}). It can be observed that the capillary height in the presence of evaporation is slightly lower than in the absence of evaporation for both the elliptic and polynomial Ansatze. Additionally, the values obtained using polynomial Ansatz deviate further from those derived using the elliptic one. The contact angle calculated with the polynomial Ansatz also differs from the prescribed contact angle of 30 degrees and is observed to be approximately equal 45 degrees.

\subsection{Comparison with an experimental data}
\label{subsec:3}

\subsubsection{Setting of the problem}
\label{subsubsec:1}
To compare our method with experimental data, we refer to the work of Polansky et. al. \cite{Polansky_experim}. In this instance, we examine a scenario involving acetone with a wall heating condition of 0.7 $\textrm{W}$. Length of computational domain is equal 0.02 $\textrm{m}$. We define the physical parameters and the thermal parameters in Tables \ref{tab:phys_comp} and \ref{tab:ther_comp}. The boundary conditions are analogical to \ref{subsec:2}.

\begin{table}[h]
\caption{Physical parameters for a comparative simulation}
\label{tab:phys_comp}     
\begin{tabular}{p{1cm}p{1.6cm}p{1.7cm}p{1.7cm}p{1.6cm}p{1.7cm}p{1.7cm}}
\hline\noalign{\smallskip}
$\Omega_\textrm{iner}$ & $r_\mathrm{cap}$, $\mathrm{m}$ & $ \rho$, $\mathrm{kg/m^3}$ & $ \mu$, $\mathrm{Pa \cdot s}$ & $g$, $\mathrm{m/s^2}$ & $\sigma$, $\mathrm{N/m}$ & $\theta_e$, $^0$ \\
\noalign{\smallskip}\hline\noalign{\smallskip}
8.41 & 0.001 & 749 & 0.0295 & 2.087 & 0.0176 & 20\\

\noalign{\smallskip}\hline\noalign{\smallskip}
\end{tabular}
\end{table}

\begin{table}[h]
\caption{Thermal parameters for a comparative simulation}
\label{tab:ther_comp}     
\begin{tabular}{p{2.0cm}p{2.0cm}p{2.0cm}p{2.0cm}p{2.0cm}p{1.1cm}p{1.1cm}}
\hline\noalign{\smallskip}
$c_\textrm{l}$ & $c_\textrm{v}$ & $ k_\textrm{l}$ & $ k_\textrm{v}$ & $h_\textrm{vap}$ & $T_\textrm{sat}$  \\
\noalign{\smallskip}\hline\noalign{\smallskip}
 $\mathrm{J / (kg \cdot K)}$ & $\mathrm{J / (kg \cdot K)}$  &  $\mathrm{W / (m \cdot K)}$ &  $\mathrm{W / (m \cdot K)}$ &  $\mathrm{J / kg}$ &  $\mathrm{K}$  \\
\noalign{\smallskip}\hline\noalign{\smallskip}
2180 & 1430 & 0.160 & 0.016 & 518  $ \cdot 10^3$ & 329.4 \\

\noalign{\smallskip}\hline\noalign{\smallskip}
\end{tabular}
\end{table}

\subsubsection{Results}
\label{subsubsec:2}
\begin{figure}[h]
  \centering
  \begin{subfigure}[b]{0.48\textwidth}
    \centering
    \includegraphics[width=\textwidth]{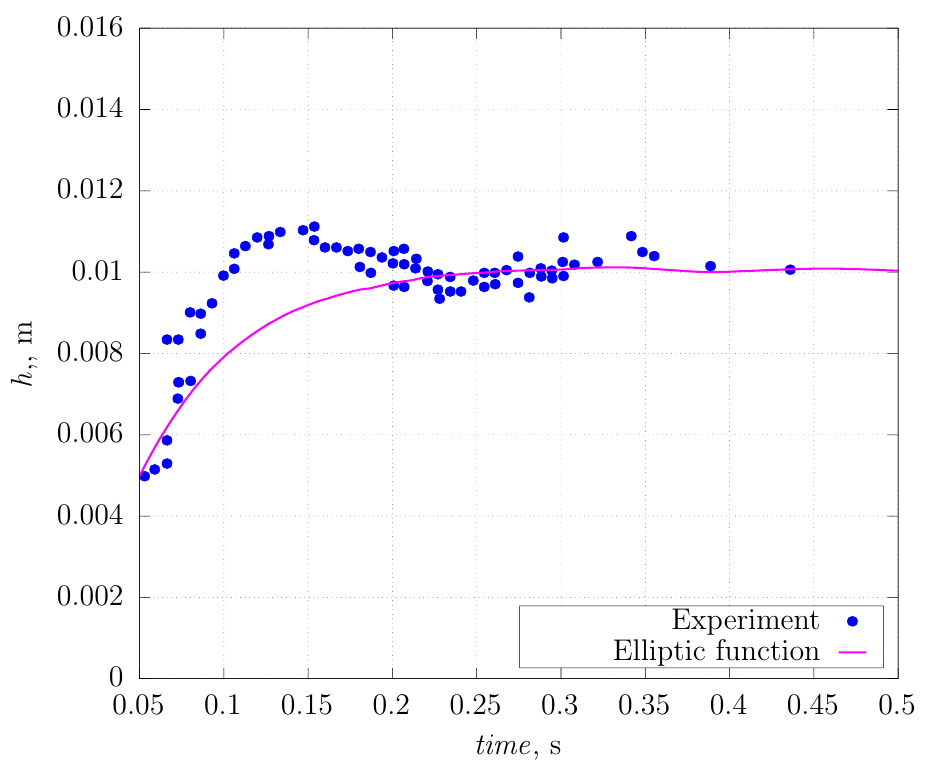}
    \caption{Capillary height}
    \label{fig:height_comp}
  \end{subfigure}
  \hfill
  \begin{subfigure}[b]{0.48\textwidth}
    \centering
    \includegraphics[width=\textwidth]{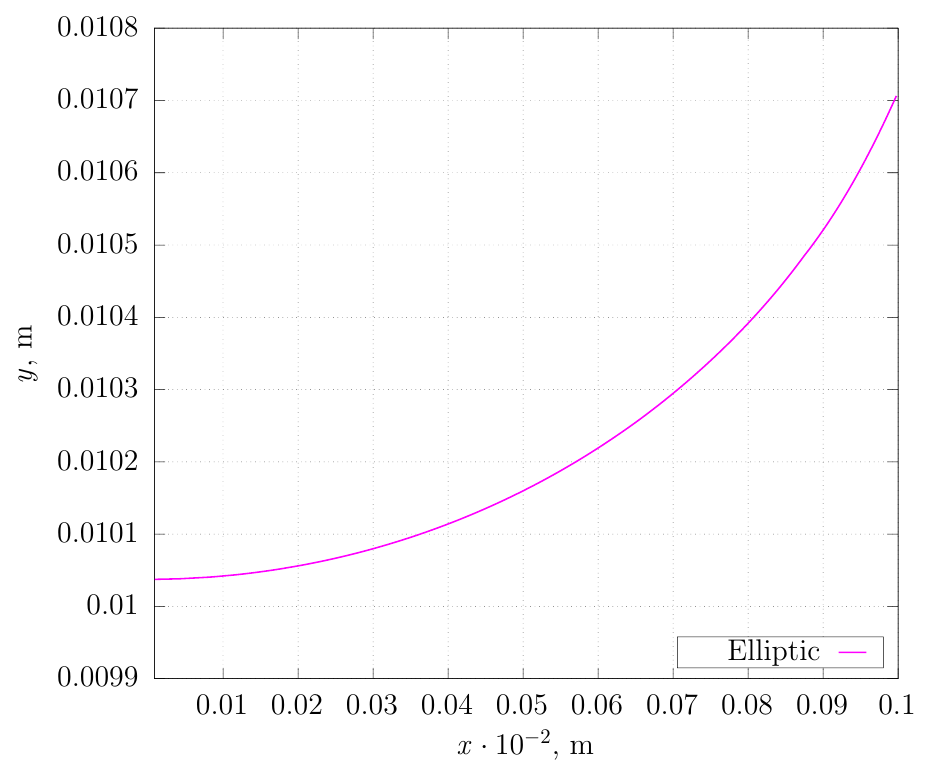}
    \caption{Interface shape}
    \label{fig:interface_comp}
  \end{subfigure}
  \caption{Capillary rise comparative simulation using elliptic Ansatz}
  \label{fig:capillary_rise_comp}
\end{figure}

As the elliptic Ansatz demonstrated better agreement with theoretical data and the results reported in the work of Gründing et. al. \cite{smuda_capillary}, we present a comparison between the experimental data and the Parameterized Level-Set Method using the elliptic Ansatz (Figure \ref{fig:height_comp}, \ref{fig:interface_comp}). The numerical results show that the capillary height values align well with the experimental data, and the contact angle closely matches the prescribed value of 20 degrees.

\section{Summary}
\label{sec:5}

The Parameterized Level-Set Method is introduced as a new approach for representing interfaces in scenarios involving the formation of a three-phase contact line. This method enhances the stability of numerical simulations and helps overcome the capillary time step restriction, thereby reducing computational costs significantly.

The simulation results using this method were compared with data from the literature. It was observed that employing an elliptic Ansatz for the level-set representation yields good agreement with theoretical values of capillary rise and the results reported in the work of Gründing et. al. \cite{smuda_capillary}. In contrast, capillary height values obtained using a polynomial Ansatz showed significant discrepancies. Similarly, contact angle values derived from the polynomial Ansatz differed from the prescribed contact angle.

Additionally, the use of the elliptic Ansatz demonstrated agreement with experimental data from Polansky et. al. \cite{Polansky_experim}.

In conclusion, the Parameterized Level-Set Method with an elliptic Ansatz is recommended as it provides excellent agreement with theoretical and experimental data while significantly reducing computational costs. This makes it an effective and efficient tool for a complicated physical scenarios with appearence of three-phase contact line.

\begin{acknowledgement}
We kindly acknowledge the financial support by the joint DFG/FWF Collaborative Research Centre CREATOR (CRC – TRR361/F90) at TU Darmstadt, TU Graz and JKU Linz.
\end{acknowledgement}
%


\bibliographystyle{plain} 
\bibliography{mybibliography} 
\end{document}